\documentstyle[aps,pra]{revtex}
\tightenlines
\newcommand{\be}{\begin{equation}}
\newcommand{\ee}{\end{equation}}
\newcommand{\ben}{\begin{eqnarray}}
\newcommand{\een}{\end{eqnarray}}
\newcommand{\bF}{\begin{figure}}
\newcommand{\eF}{\end{figure}}
\title{Comments on Struyve and Baere's paper on
experiments to distinguish Bohmian mechanics from quantum
mechanics}
\author{Partha Ghose}
\address{S. N. Bose National Centre for Basic Sciences,
Block JD, Sector III, Salt Lake, Kolkata 700 098}
\begin{document}
\maketitle

\begin{abstract} It is shown in detail why the arguments put
forward by Struyve and Baere (quant-ph/0108038) against my
conclusions are incorrect.
\end{abstract}
\vskip 0.2in There are several strands to the arguments I used
that seem to have been overlooked by the authors. Let me take them
up one by one. But right in the beginning I would like to
emphasize that two aspects of the discussion must be clearly
separated from each other, the question of (a) the theoretical
incompatibility between dBB and quantum mechanics, a question that
can be settled with the help of {\it gedanken} experiments alone,
and (b) the feasibility of real experiments to test this
incompatibility.

\vskip 0.2in
\noindent{\it Quantum Equilibrium Hypothesis}
\vskip 0.2in
Without loss of generality let us restrict the discussion to the
two-particle case. In dBB particle trajectories ${\bf x}_1(t)$ and
${\bf x}_2(t)$ are introduced through the {\it guidance condition}
\be
{\bf p}_{i} =   \nabla_{{\bf x}_i} S({\bf x}_1,{\bf x}_2,
t)\label{eq.1}\ee where $S({\bf x}_1,{\bf x}_2, t)$ is the phase
of the wave function $\psi ({\bf x}_1,{\bf x}_2, t)$ in quantum
mechanics. (I use the notation ${\bf x} = (x, y, z)$.) Having
introduced the trajectories, one then postulates that
\be
\rho({\bf x}_1(t_0),{\bf x}_2(t_0)) = \vert \psi ({\bf x}_1,{\bf
x}_2, t_0) \vert^2\label{eq.2} \ee at some time time $t_0$. Then
the continuity equation guarantees that this relation is valid at
all times $t$:
\be
\rho({\bf x}_1(t),{\bf x}_2(t)) = \vert \psi ({\bf x}_1,{\bf x}_2,
t) \vert^2\label{eq.3} \ee This is the only hypothesis in dBB that
may be called the quantum equilibrium hypothesis (QEH). Writing it
in the form
\be
\rho({\bf x}_1,{\bf x}_2, t) = \vert \psi ({\bf x}_1,{\bf x}_2, t)
\vert^2\label{eq.4} \ee is misleading because this relation is a
mathematical identity without any physical content. The guidance
condition being the only input in dBB that is additional to
quantum mechanics, it goes without saying that any analysis that
does not take this condition into account and merely uses the
identity (\ref{eq.4}) and the continuity equation will trivially
reproduce the same results as quantum mechanics in every case.

Note that ${\bf x}_1$ and ${\bf x}_2$ are variables that cover the
entire support of the wave function $\psi$, and hence the
trajectories ${\bf x}_1(t)$ and ${\bf x}_2(t))$ must also cover
the same support. Therefore, to realize the distribution
$\rho({\bf x}_1(t),{\bf x}_2(t))$ one must have an infinitely
large number of particles of types $1$ and $2$ distributed in the
prescribed manner $\rho$ at time $t$. One then has a full
ensemble.

If we take any observable $\cal{O}$, its (space) average in dBB is
then given by

\be
\bar{O} = \int_t {\cal{O}} \rho({\bf x}_1(t),{\bf x}_2(t))\, d{\bf
x}_1(t) d{\bf x}_2(t)\label{eq.5} \ee This tells us that one has
to integrate over {\it all possible trajectories ${\bf x}_1(t)$
and ${\bf x}_2(t)$} {\it at a fixed time} $t$ in order to compute
the average in dBB. This is the ensemble average.

Now consider a theoretically admissible situation in which one has
particle $1$ at a definite position ${\bf X}_1(t)$ and particle
$2$ at ${\bf X}_2(t)$ at some time $t$ and {\it everywhere else
the wave function (in dBB) is empty}. Since, {\it by hypothesis,
empty parts of the wave function do not fire detectors}, only two
trajectories out of all the possible ones contribute to detections
in this case. If one repeats the situation over and over again
with identical copies of the system at different times $(t_1, ...,
t_i, ..., t_N)$, the particles will be in different positions
${\bf X}_1(t_i)$ and ${\bf X}_2(t_i)$ each time with probabilities
determined by the distribution function $\rho({\bf X}_1(t_i),{\bf
X}_2(t_i))$. If one collects all the results together and computes
the time average of the joint detection probability, one would
obtain (for two detectors $D_1$ and $D_2$ placed with their $x$
and $z$ coordinates the same and fixed)

\ben P^*_{12} &=& lim_{N \rightarrow \infty}
\frac{1}{N}\sum_{t_1}^{t_N} \int_{D_1,D_2,t_i} \rho({\bf
x}_1(t_i),{\bf x}_2(t_i))\, \delta ({\bf x}_1(t_i) - {\bf
X}_1(t_i))\, \delta ({\bf x}_2(t_i) - {\bf X}_2(t_i)) \, d
y_1(t_i)\, d y_2(t_i)\\&=& lim_{N \rightarrow \infty}
\frac{1}{N}\sum_{t_1}^{t_N} \rho({\bf X}_1(t_i),{\bf X}_2(t_i)) =
\bar{P}_{12}\label{eq.6} \een where $\bar{P}_{12}$ is the space
average. This result is, however, true only provided {\it the
guidance condition does not imply a constraint like $Y_1(t_i) +
Y_2(t_i) = 0$ on the trajectories for every $t_i$}.

If there is a constraint like $\delta(Y_1(t_i)+ Y_2(t_i))$ on the
trajectories due to some symmetry, it will make the system
non-ergodic. One would then obtain

\be P^*_{12} = lim_{N \rightarrow \infty}
\frac{1}{N}\sum_{t_1}^{t_N}\frac{1}{\delta(0)} \int_{D_1,D_2,t_i}
\rho({\bf x}_1(t_i),{\bf x}_2(t_i))\, \delta ({\bf x}_1(t_i) -
{\bf X}_1(t_i))\, \delta ({\bf x}_2(t_i) - {\bf
X}_2(t_i))\nonumber\ee \be .\,\,\delta(Y_1(t_i)+ Y_2(t_i))\, d y
_1(t_i)\, d y_2(t_i) \neq \bar{P}_{12}\label{eq.7}\ee
Nevertheless, if one computes the time average of an observable
other than the joint detection probability, one would still obtain

\be O^*_{12} = lim_{N \rightarrow \infty}
\frac{1}{N}\sum_{t_1}^{t_N}\frac{1}{\delta(0)} \int_{t_i}
{\cal{O}} \rho({\bf x}_1(t_i),{\bf x}_2(t_i))\, \delta ({\bf
x}_1(t_i) - {\bf X}_1(t_i))\, \delta ({\bf x}_2(t_i) - {\bf
X}_2(t_i))\nonumber\ee \be .\,\,\delta(Y_1(t_i)+ Y_2(t_i))\, d y
_1(t_i)\, d y_2(t_i) = \bar{O}\label{eq.8} \ee because the
constraint, being a consequence of the guidance condition, is
necessarily consistent with the distribution function $\rho$ by
hypothesis, and most importantly {\it the restrictions on the
integration variables coming from the supports of the detectors is
removed}.

All this shows that QEH is applicable to the full ensemble but not
to individual processes that make up this ensemble. There could be
information regarding individual processes determined by Bohmian
dynamics (through the guidance condition) that are hidden in the
full ensemble. The symmetric trajectories in a double-slit
experiment with single particles is a clear example of how the
information that the trajectories are symmetrical about the line
of symmetry and do not cross, is masked in the distribution
function $\rho({\bf x}(t))$.

With this background let us now consider a {\it gedanken}
experiment in which two particles are incident at a time on a
screen $S$ with two point-like slits (the points being assumed to
be limits of spheres). The source of the particles is assumed to
be such that it produces entangled pairs of particles, one pair at
a time, and it is so placed that out of all the pairs it emits, a
significant number is produced such that one particle passes
through each slit. The rest of the pairs are blocked by the screen
$S$. On the far side of the slit two spherical waves will emerge
from the two slits. I have shown by an explicit analytical
calculation (quant-ph/0103126) that the trajectories of the
particles will be symmetrical about the line of symmetry between
the slits. The system is therefore non-ergodic. Let two detectors
be placed at a fixed distance $X$ along the $x$-axis
asymmetrically about the line of symmetry $x=0$. It follows from
eqn. (8) that $P^*_{12} = 0$ in this case. Since $\bar{P}_{12}
\neq 0$, this establishes the incompatibility.

If one widens the slits somewhat and uses Gaussian profiles, then,
as Struyve and Baere have shown,
\be
\sigma(t) \equiv Y_1(t) + Y_2(t) = \sigma(0) \sqrt{1 +
(\hbar/2m\sigma_0^2)^2 t^2)} \label{eq.9}\ee This spreading can be
made as small as one desires by requiring that
\be
(\hbar/2m\sigma_0^2)^2 t^2 << 1 \ee There is nothing in the theory
that precludes such a condition. Since this spreading is
calculable in any case, it can be taken care of in the design of
the experiment.

\vskip 0.2in \noindent{\it Ergodicity} \vskip 0.2in First of all,
let me state that the definition (and proof) of ergodicity of
quantum mechanics that I gave in quant-ph/0103126 is not due to me
but can be found in standard text books such as that by Toda {\it
et al}\cite{toda}. Furthermore, the cross terms have not been
ignored in the proof, as the authors seem to imply, but they can
actually be shown to vanish in the limit of large time. In fact,
this is the most crucial part of the proof supplied by
mathematicians in the most general case. Having said that, let me
also state that the proof of ergodicity of dBB given by Struyve
and Baere is flawed by the trivial use of QEH in the form
(\ref{eq.4}) as explained above.

Finally, let me comment on their equation (26),
\be
\int_M f^* d \mu = \bar{f} \ee What the authors have not stated is
the condition under which this theorem is valid, namely, that the
flow be such that there is no invariant subspace of the full phase
space $M$ , i.e., $\{\phi_t(E)= E \subseteq M \Rightarrow E \equiv
M\,\, {\rm or}\,\, \emptyset \}$. In other words, this theorem is
valid for ergodic systems only. Therefore there is no
inconsistency in my arguments.

\vskip 0.2in \noindent{\it Double pendulum} \vskip 0.2in In this
case it is obvious that the Bohmian motion of the individual
pendulums remains periodic (as in the classical case which is
non-ergodic if the two eigenfrequencies are commensurate), and
therefore the motion does not cover the entire phase space torus
That is sufficient to prove non-ergodicity for commensurate
frequencies. Since there is a general theorem that states that
there must be at least one observable in a non-ergodic system
whose space and time averages are different, the incompatibility
with quantum mechanics is established.

\end{document}